\documentclass[aps,pre,twocolumn,floatfix,superscriptaddress]{revtex4}
\usepackage{graphicx}
\usepackage[colorlinks=true,linkcolor=blue,citecolor=blue,urlcolor=blue]{hyperref}
\usepackage{xcolor}
\usepackage{etoolbox}
\usepackage{tikz}
\usepackage{amsmath}
\usepackage[capitalize]{cleveref}
\newrobustcmd*{\mycircle}[1]{\tikz{\filldraw[draw=#1,fill=#1] (0,0) circle [radius=0.1cm];}}
\definecolor{col1}{rgb}{0.7 0 0}
\definecolor{col2}{rgb}{0 0.7 0}
\definecolor{col3}{rgb}{0 0 0.7}

\begin{document}
\makeatletter

\title{The following article has been accepted by Chaos, AIP\\ Explosive synchronization in temporal networks: A comparative study}

\author{Tanu Singla}
\email[Corresponding author:~]{tanu.singla@tec.mx}
\affiliation{Tecnol\'ogico de Monterrey, Calle del Puente 222, Colonia Ejidos de Huipulco, Tlalpan, Ciudad de M\'exico, M\'{e}xico - 14380}

\author{M. Rivera}
\affiliation{Centro de Investigaci\'on en Ciencias-(IICBA), UAEM, Avenida Universidad 1001, Colonia Chamilpa, Cuernavaca, Morelos, M\'{e}xico - 62209}

\date{\today}

\begin{abstract}

We present a comparative study on Explosive Synchronization (ES) in temporal networks consisting of phase oscillators. The temporal nature of the networks is modeled with two configurations: (1) oscillators are allowed to move in a closed two dimensional box such that they couple with their neighbors, (2) oscillators are static and they randomly switch their coupling partners. Configuration (1) is further studied under two possible scenarios: in the first case oscillators couple to fixed numbers of neighbors while in other they couple to all oscillators lying in their circle of vision. Under these circumstances, we monitor the degrees of temporal networks, velocities, and radius of circle of vision of the oscillators, and the probability of forming connections in order to study and compare the critical values of the coupling required to induce ES in the population of phase oscillators.

\end{abstract}

\maketitle

{\bf ES is a phenomenon in which a first order transition in the extent of synchronization is observed as the coupling strength between oscillators increases. In this work, we report ES in networks of phase oscillators where the coupling partners of the oscillators change during time. Three different configurations are used to implement temporal nature of the networks. Under these conditions, we compare our results (for three configurations) to observe ES in the population of coupled phase oscillators.}

\section{Introduction}
Synchronization is a phenomenon wherein simultaneous evolution in the dynamics of oscillators is observed by the virtue of coupling between them. It was discovered by Christian Huygens when he performed experiments with a pair of pendulum clocks hanging on a common supporting beam. Although several studies on synchronization were reported after Christian Huygens, the work on phase synchronization in chaotic oscillators by Rosenblum et al. in 1996 \cite{Rosenblum} motivated scientists to explore synchronization in different systems. Since then, synchronization and its manifestations have been studied in diverse systems found in nature \cite{Pikovsky,Strogatz,Stern,Dotson} as well as in model experimental \cite{Rita,Tanu,Jyoti, Nguyen} and numerical systems \cite{Prasad,Stefa,Chowdhury}.

In the pioneering work, Y. Kuramoto \cite{Kuramoto1} studied synchronization in a network of globally coupled phase oscillators and a second order transition from unsynchronized to synchronized state was reported. This was followed by a plethora of studies for synchronization in networks involving phase oscillators and other dynamical systems. Chimera state is a manifestation of synchronization in networks where due to the nature of coupling, a partial population of the oscillators synchronizes and the other oscillators remain unsynchronized \cite{Kuramoto,Abrams}. Synchronization has also been explored in random networks wherein the oscillators randomly couple to each other \cite{Arenas}. Finally, synchronization has also been studied in temporal networks. The topology of temporal networks changes over time {\em i.e.} the oscillators keep forming new connections with other oscillators while losing and/or maintaining some of the old connections. The investigation of temporal networks has its relevance in understanding functional brain networks \cite{Ramirez}, power transmission systems \cite{Ji}, epidemic spreading \cite{Ishant}, robotics, etc. In nature, temporal networks can be observed in a population of fireflies where a firefly observes the blinking of its neighbors and adjust its frequency of blinking. The adjustment of rhythms by the entire population of fireflies results in the emergence of synchronized behavior. Two different variations of temporal networks are studied and reported: in one case, the oscillators remain static while switching couplings with other oscillators \cite{Amritkar,Kohar,Rakshit} and in the other case, the oscillators move in a given space and they couple to their nearby oscillators \cite{Sumantra,Lavneet,Prignano,Levis}. Most of these works on temporal networks (static or moving oscillators) reported synchronization, its stability, second order Kuramoto transition, etc. However, up to the best of our knowledge, ES has not yet been explored in temporal networks.

ES is a manifestation of synchronization in a population of interacting oscillators which is characterized by a first order transition from unsynchronized to synchronized state when the strength of interaction ($\epsilon$) between the oscillators is gradually increased. The presence of hysteresis is another property that is associated with the first order transition in ES. This signifies that the value of coupling strength ($\epsilon=\epsilon_l$) at which oscillators unsynchronize while decreasing coupling between oscillators is lower than the critical value of coupling strength ($\epsilon=\epsilon_u$) to achieve synchronization while increasing coupling. The principal mechanism behind this phenomenon is to suppress the formation of synchronization clusters which eventually results in the global synchronization of the population. In the initial works on ES, it was shown that this suppression can be achieved by implementing a positive correlation between the natural frequencies of the oscillators and their degrees \cite{Gomez,Leyva}. Following these articles, several other mechanisms to induce first order transition have been reported \cite{Leyva1,Peng,Amit,Boccaletti}. In another report by Zhang et al. \cite{Zhang}, the role of local order parameter of the oscillators in achieving ES has been explored by implementing its adaptive control on the dynamics of the oscillators. This adaptive control reduces the effective coupling between oscillators as the local order parameter starts to augment, which further causes the suppression of synchronization clusters. Eventually, when the coupling strength between oscillators is sufficiently high (higher than critical $\epsilon$ to observe second order Kuramoto transition), ES can be observed in the population of oscillators. In another work \cite{Kanhra} this concept has been extended to modify the width of the hysteresis loop formed during ES. The mechanism of ES has also been used to explain a neurological disease called Fibromyalgia \cite{Lee}. A functional network from the EEG signals of the patients was constructed and analyzed to identify the imprints of ES. Finally, ES has also been reported in a model experimental system consisting of Mercury Beating Heart oscillators \cite{Pawan}.

In this work, we present a comparative study of ES in temporal networks. Three different configurations are considered to employ temporal coupling in a population of phase oscillators. In one of the configurations, the oscillators remain static and randomly couple to other oscillators such that the average degree of the networks forming at any time moment fluctuates around a mean value. In the other two configurations, the oscillators execute random walk in a two dimensional closed space and couple to their neighbors; degrees of the oscillators for these cases are further decided on the basis of two different schemes (discussed later). Depending on the configuration, parameters like coupling strength, velocity, vision size of oscillators, and the probability of forming connections between oscillators are varied to study ES.

\section{Coupling Mechanisms}

We consider a population of $N$ coupled phase oscillators to study ES in temporal networks. A population of coupled phase oscillators is generally represented with the following equation:

\begin{equation}
\dfrac{d \phi_p}{dt} = \omega_p+\frac{\epsilon}{k_p} \sum_{q=1}^N A_{pq}\sin(\phi_q-\phi_p).
\label{eq1}
\end{equation}
Here, $\phi_p$ and $\omega_p$ are the phase and the frequency of the $p^{th}$ oscillator, $\epsilon$ is the coupling strength between the oscillators, and $A_{pq}$ is an element of adjacency matrix $A$ giving details about the coupling links between the oscillators; $A_{pq}=1$ if $p^{th}$ and $q^{th}$ are coupled and $A_{pq}=0$ otherwise. Degree of an oscillator can be calculated from $k_p=\sum_{q=1}^N A_{pq}$. The extent of synchronization (instantaneous order parameter) of the population can be calculated using the following expression:

\begin{equation}
re^{i\Phi} = \frac{1}{N}\sum_{q=1}^N e^{i\phi_q}.
\label{eq2}
\end{equation}
Here, $0\leq r\leq1$ represents the extent of synchronization and $\Phi$ represents the average phase of the population.

Our scheme of implementing coupling in phase oscillators is a combination of the schemes proposed by \cite{Zhang}, and \cite{Kanhra}, where the dynamics of the oscillators were influenced with their local/global order parameters. Along with employing order parameter dependence, we also normalized the coupling function with the degree ($k_p$) of the oscillators; as studied in \cite{Motter}, normalizing the coupling function with weights is essential to synchronize a heterogeneous network. Under such circumstances, the model equation of the coupled population that we use is:

\begin{equation}
\dfrac{d \phi_p}{dt} = \omega_p+\frac{\epsilon r_p^2}{k_p}\sum_{q=1}^N A_{pq}\sin(\phi_q-\phi_p).
\label{eq3}
\end{equation}
Similar to \cref{eq2}, the instantaneous local order parameter of an oscillator in our model is defined as:

\begin{equation}
r_pe^{i\Phi_p} = \frac{1}{k_p}\sum_{q=1}^N A_{pq}e^{i\phi_q},
\label{eq4}
\end{equation}
$\Phi_p$ is the average phase of the oscillators coupled to $p^{th}$ oscillator. Frequencies of the oscillators ($\omega$) are chosen from a uniform distribution of numbers lying between interval (0,2]. Finally, to simulate temporal networks the instantaneous adjacency matrices $A_{pq}$ are modeled using following three configurations:

\begin{itemize}
\item {\bf Configuration 1:} Moving oscillators with nearest neighbor coupling such that every oscillator of the population couple to same number of nearest oscillators {\em i.e.} $k$ of every oscillator remains equal and constant.

\item {\bf Configuration 2:} Moving oscillators with nearest neighbor coupling such that every oscillator has a circular vision size and the oscillator couple to all oscillators lying in its vision. In this case, $k$ depends on the radius ($R$) of the circle of vision.

\item {\bf Configuration 3:} Static oscillators where every oscillator randomly couples to other oscillators such that the average degree of the temporal networks remains the same throughout time. In this case, $k$ depends on the probability ($p$) with which oscillators couple with each other.
\end{itemize}

\begin{figure}[ht!]
\includegraphics[width=9cm,height=7.5cm]{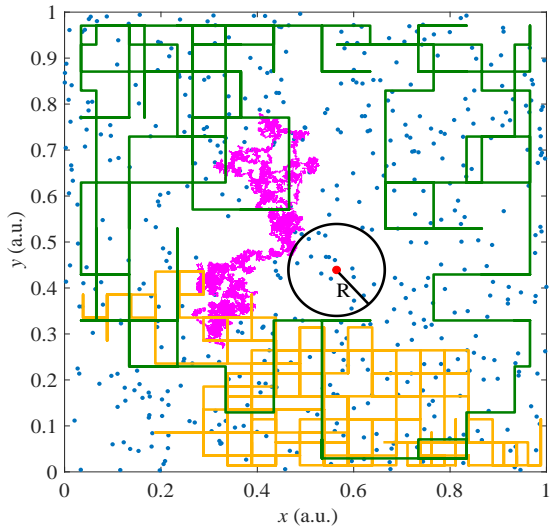}
\caption{Demonstration of instantaneous locations of $N=500$ phase oscillators in the $x-y$ plane along with the trajectories of oscillators with $v=0.002$ (magenta), $v=0.05$ (yellow), and $v=0.1$ (green) curves and also the circle of vision (black) of one of the oscillators (red dot) with $R=0.1$. For the purpose of demonstration, only few steps have been shown for oscillators moving with $v=0.05$, and $v=0.1$.}
\label{1}
\end{figure}

\begin{figure*}[ht!]
\includegraphics[width=18cm,height=10cm]{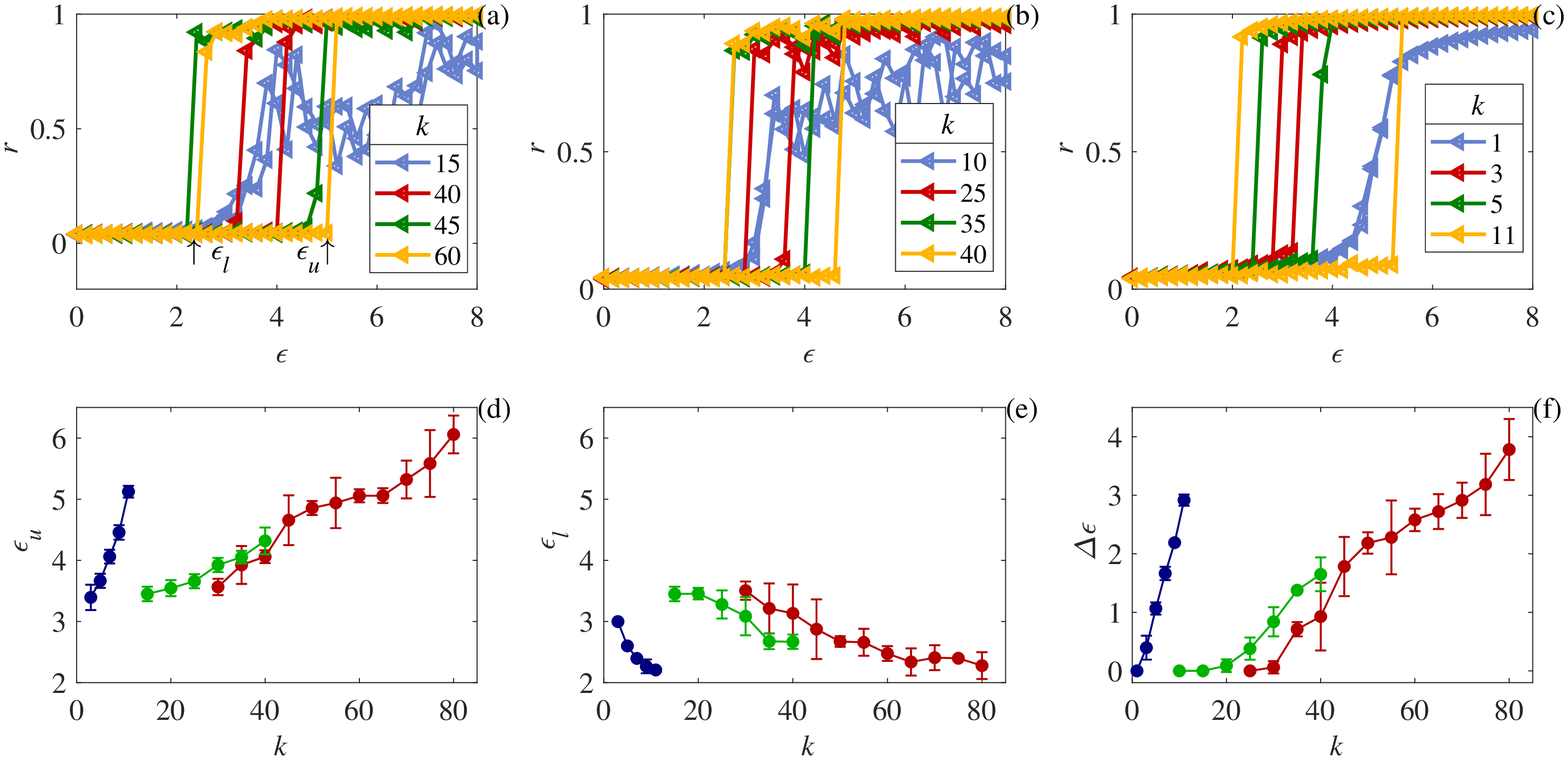}
\caption{(a-c) Results of ES in temporal networks emulated with configuration 1 for three different velocities of the oscillators: (a) $v=0.002$, (b) $v=0.01$, (c) $v=0.1$. (d) Critical values of the coupling strength required to synchronize oscillators, (e) critical values of coupling strength required to unsynchronize oscillators, and (f) width of the hysteresis loops formed during ES. \mycircle{col1} ($v=0.002$), \mycircle{col2} ($v=0.01$), and \mycircle{col3} ($v=0.1$).}
\label{2}
\end{figure*}

In the case of configurations involving moving oscillators, their motion is confined into a unit size two dimensional box with rigid boundaries $i.e$ oscillators reflect back into the box at the boundaries. Oscillators are permitted to move a $\delta$ step in either $\pm x$ or $\pm y$ direction at every iteration. The value of the step moved by oscillators also defines their velocity ($v=\delta$) in the closed box. In \cref{1} the initial positions (blue dots) of $N=500$ oscillators are shown for demonstration. The motion of three random oscillators with different velocities is represented with magenta line ($v=0.002$), yellow line ($v=0.05$), and green line ($v=0.1$). A circle of vision of an oscillator (marked with a red dot) belonging to configuration 2 with radius of vision $R=0.1$ is also shown with a black circle. Further details on the temporal aspects of networks for each configuration are discussed later when their respective results are presented. Finally, \cref{eq3} is numerically simulated using the RK4 algorithm with a step size of $dt=0.02$ and 25000 iterations; in \cite{Odor} authors report that $dt=0.1$ is sufficiently small to numerically simulate a population of coupled phase oscillators. Results for $\epsilon_u$, $\epsilon_l$ and $\Delta\epsilon$ are obtained by performing three simulation runs at different parameters and their average values along with the associated standard deviations are presented on different figures.

\section{Results}

\subsection{Configuration 1}

In this configuration, the temporal nature of the networks of phase oscillators is such that the oscillators move in a closed unit size box with uniform velocities and all of the oscillators interact with a fixed number of nearest neighbors. This implies that the instantaneous degree ($k$) of all oscillators remains the same and fixed. Moreover, velocities ($v$) of oscillators are equal to the step $\delta$ that the oscillators move in every iteration.

\cref{2} shows the results of ES in phase oscillators modeled with this configuration. In \cref{2}(a-c), we plot the variation of order parameter of the entire population as a function $\epsilon$ for three different velocities of the oscillators ((a): $v=0.002$, (b) $v=0.01$, and (c) $v=0.1$). Moreover, for each of these cases, the order parameter is calculated and plotted for various values of $k$. It can be noted that for a sufficiently small degree of the oscillators ($k=15$ and $v=0.002$ in \cref{2}(a)), the population of the oscillators undergoes classical second order Kuramoto transition in the order parameter. However, as $k$ increases gradually, the population experiences explosive (first order) transitions between unsynchronized and synchronized states; the transition is also accompanied by its characteristic hysteresis loop. For the purpose of demonstration, the critical values of coupling strength at which oscillators synchronize (unsynchronize) are marked with $\epsilon_u$ ($\epsilon_l$) on \cref{2}(a). Furthermore, similar dependences of ES on $k$ are also observed when the oscillators move with higher velocities: $v=0.05$ and $v=0.1$ (\cref{2}(b and c)).

In \cref{2}(d), the variation of $\epsilon_u$ (critical value of coupling required to observe ES) is illustrated as a function of $k$ for three different velocities; for any $v$, the lowest value of $k$ is the degree of the oscillators at which ES starts to appear. It can be noted that $\epsilon_u$ increases uniformly with $k$ for every velocity. In \cite{Restrepo,Rodrigues} Kuramoto model has been studied by the perspective of complex networks. According to this study, if the time averaged local order parameter of an oscillator in the present situation is defined with the following relation:

\begin{equation}
r_p^\prime e^{i\Phi_p}=\sum_{q=1}^N A_{pq}\langle e^{i\phi_q}\rangle_t,
\label{eq5}
\end{equation}
where $\langle\cdots\rangle_t$ is the time average, then, the state of an oscillator can be represented as:

\begin{equation}
\dfrac{d \phi_p}{dt} = \omega_p+\frac{\epsilon r_p^{2}r_p^\prime}{k_p}\sin(\Phi_p-\phi_p)-\epsilon h_p.
\label{eq6}
\end{equation}
Here, $\Phi_p$ is the average phase of the oscillators coupled to $p^{th}$ oscillator and $h_p$ accounts for the time fluctuations in the dynamics of this oscillator by the virtue of temporal adjacency matrices. The mathematical form of $h_p$ is given by $h_p=\mathrm{Im}\{e^{-i\phi_p}\sum_qA_{pq}(\langle e^{i\phi_q}\rangle_t-e^{i\phi_q})\}$, where ``Im" stands for imaginary and $\langle\cdots\rangle_t$ is the time average. During the onset of synchronization, $r_p=r_p^\prime\mathtt{\sim}k_p$ and $h_p$ is expected to be of the order of $\sqrt{k_p}$ ($k_p$ is the degree of $p^{th}$ oscillator). Therefore, as the degree of the oscillators of the moving population increases, these fluctuations also increase causing the suppression of synchronization clusters (as mentioned in the introduction section) and requiring even larger $\epsilon$ for the population to synchronize. Another interesting behavior that can be observed in \cref{2}(d) is that the minimum value of $k$ at which ES starts to appear reduces as the velocities of the oscillators increase. The instantaneous time spent by an oscillator in a local cluster decreases as its velocity increase, causing less interaction among the members of the clusters. The results of \cref{2}(d) show that when the degree of oscillators is large then due to higher heterogeneity in the cluster, it would require more time or higher coupling strength to synchronize at higher velocities. Conversely, it can be said that as the velocity of the oscillators increases, the population of the oscillators can exhibit ES at lower degrees. In \cref{2}(e) variation of $\epsilon_l$ (critical value of coupling at which oscillators unsynchronize) as a function of $k$ for three different velocities is shown and it can be observed that $\epsilon_l$ decreases with $k$ for different velocities. Furthermore, the critical values of $\epsilon_u$ and $\epsilon_l$ constitute the hysteresis loops of ES and in agreement to the results of \cref{2}(d and e), the width of the hysteresis loops increases with $k$ (\cref{2}(f)).

\begin{figure}[ht!]
\includegraphics[width=9cm,height=8cm]{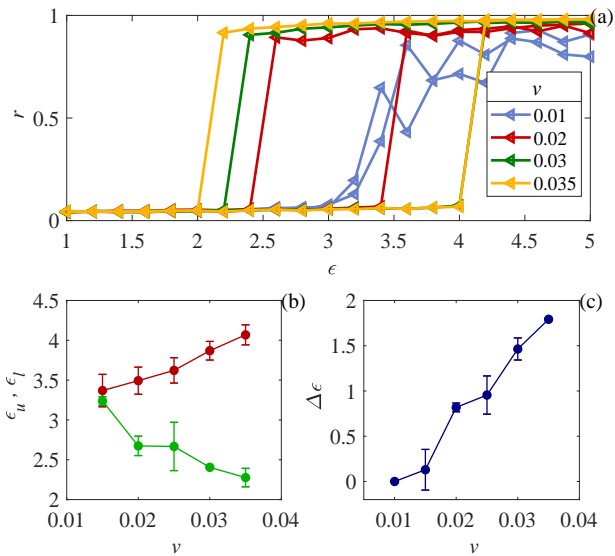}
\caption{(a) Results of ES in temporal networks emulated with configuration 1 for $k=15$. (b) Critical values of the coupling strength required to synchronize ($\epsilon_u$: \mycircle{col1}), and critical values of coupling strength required to unsynchronize oscillators ($\epsilon_l$: \mycircle{col2}), (c) width of the hysteresis loops formed during ES.}
\label{3}
\end{figure}

From \cref{2}(a) it can be observed that when $k=15$, the population of oscillators does not exhibit ES for $v=0.002$ but it appears at higher velocities for the same value of $k$ (\cref{2}(b and c)). To study this transition, we vary velocities of the oscillators keeping $k=15$ and the results of explosive transitions are shown in \cref{3}(a). It can be observed that the oscillators exhibit ES as their velocities increase from $v=0.01$ to $v=0.02$. Furthermore, in \cref{3}(b), variation of $\epsilon_u$ is shown as a function of $v$. In \cite{Lavneet} it has been reported that similar to coupling strength ($\epsilon$), the velocity of the oscillators also acts as a parameter to observe second order Kuramoto transition of synchronization. In the present case, when velocities of the oscillators augment, it results in the formation of synchronization clusters. From the analysis of the dependence of $\epsilon_u$ on $k$ (\cref{eq6}), it can be said that the local order parameter of the oscillators also depends on the velocities of the oscillators and it explains the increasing nature of $\epsilon_u$ with $v$ (\cref{3}(b)). Finally, variations of $\epsilon_l$ and the width of the hysteresis loops are shown on \cref{3}(b and c) and they are identical to their respective results in \cref{2}(e and f).

\subsection{Configuration 2}

In this section, the results of ES in temporal networks obtained using configuration 1 are compared with another configuration in which coupling is implemented such that every moving oscillator interacts with other oscillators lying in its vision circle (radius: $R$). As a consequence, the degree of oscillators ($k$) in this case depends on $R$, does not remain fixed in time, and could be different for every oscillator. Similar to the previous subsection, results are obtained by varying $R$, keeping $v$ constant and vice versa.

\begin{figure}[ht!]
\includegraphics[width=9cm,height=8cm]{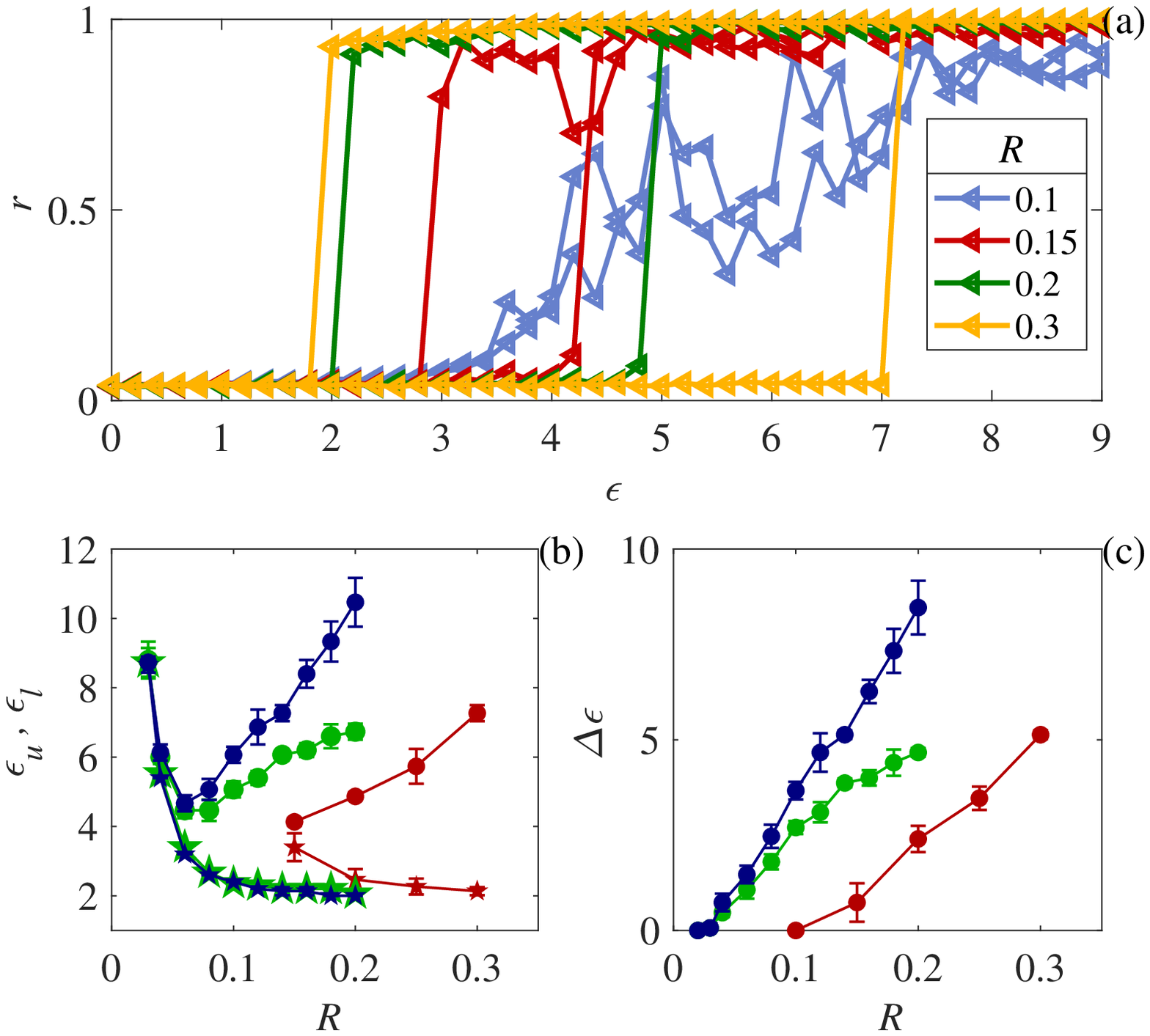}
\caption{(a) Results of ES in temporal networks emulated with configuration 2 for $v=0.002$. (b) Critical values of the coupling strength required to synchronize ($\epsilon_u$: circles), and critical values of coupling strength required to unsynchronize oscillators ($\epsilon_l$: stars), (c) width of the hysteresis loops formed during ES. \mycircle{col1} ($v=0.002$), \mycircle{col2} ($v=0.05$), and \mycircle{col3} ($v=0.1$).}
\label{4}
\end{figure}

In \cref{4}(a), hysteresis loops for different values of $R$ are shown for $v=0.002$. It can be observed that for smaller $R$, ES is not observed and that it starts to appear as $R$ increases. The variation of $\epsilon_u$ and $\epsilon_l$ as a function of $R$ is shown in \cref{4}(b) for $v=0.002$, $v=0.05$, and $v=0.1$. It must be noted that, while $\epsilon_u$ for $v=0.002$ and $\epsilon_l$ for all velocities show similar trend as their counterparts in \cref{2}(d and e), the response of $\epsilon_u$ for $v=0.05$ and for $v=0.1$, however, is different. It shows that the critical value of $\epsilon$ to achieve synchronization of oscillators moving with larger velocities shows an initial fall before starting to increase with $R$. Given that \cref{eq6} explains the relationship between $\epsilon_u$ and $k$, this result deviates for smaller $R$ ($k\propto R^2$) from its analogous results of \cref{2}(d) where the critical $\epsilon_u$ for ES uniformly increased with $k$. The possible reason for this deviation lies in the fact that in the present case $k$ of oscillators is not constant. However, detailed theoretical and/or numerical analysis needs to be carried out to understand this behavior. Finally, in \cref{4}(c), the variation in the width of hysteresis loops as a function of $R$ is plotted for different velocities and it can be observed that the width increases with $R$.

\begin{figure}[ht!]
\includegraphics[width=9cm,height=8cm]{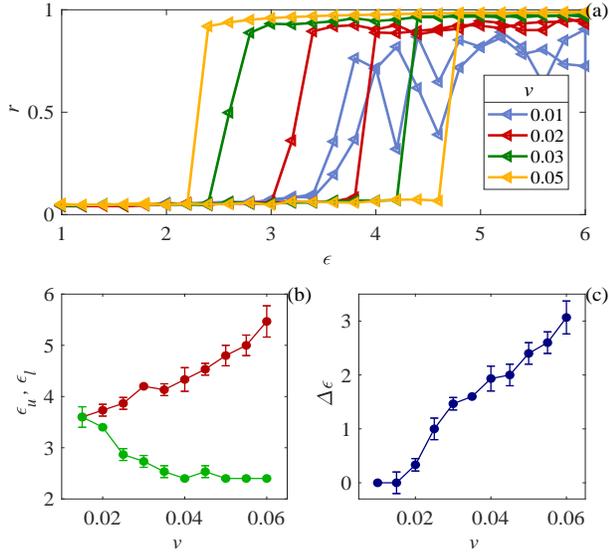}
\caption{(a) Results of ES in temporal networks emulated with configuration 2 for $R=0.1$. (b) Critical values of the coupling strength required to synchronize ($\epsilon_u$: \mycircle{col1}), and critical values of coupling strength required to unsynchronize oscillators ($\epsilon_l$: \mycircle{col2}), (c) width of the hysteresis loops formed during ES.}
\label{5}
\end{figure}

For the purpose of completion, in \cref{5}, the results of ES of moving oscillators are shown by varying $\epsilon$ and $v$; keeping $R$ of the oscillators fixed at $0.1$. Results obtained in this case are identical to those presented in \cref{3}, where it was shown that the oscillators do not exhibit ES when they move slowly and at higher velocities width of the hysteresis loops of the ES increases as the velocities of the oscillators increase.

To calculate approximate interaction time between moving oscillators, the timescales of velocities and frequencies of the oscillators is obtained by using the range of radii ($R$), velocities ($v$) and frequencies ($\omega$) for different cases. From the results in \cref{4} and \cref{5}, range of radii of the oscillators is $R=[0.02,0.3]$ and that for velocities is $v=[0.002,0.06]$. Then for two oscillators which are very close to each other, the range of ``minimum time'' that will they take so that they are out of each other's circle of vision is given by $t_v=[\frac{R_{min}dt}{2v_{max}},\frac{R_{max}dt}{2v_{min}}]=[0.0033,1.5]$. Moreover, the range of timescales related to frequencies of the oscillators is $t_f=[\frac{2\pi}{\omega_{max}},\frac{2\pi}{\omega_{min}}]=[3.14,\infty)$ ($\omega_p=(0,2]$).

\begin{figure}[ht!]
\includegraphics[width=9cm,height=8cm]{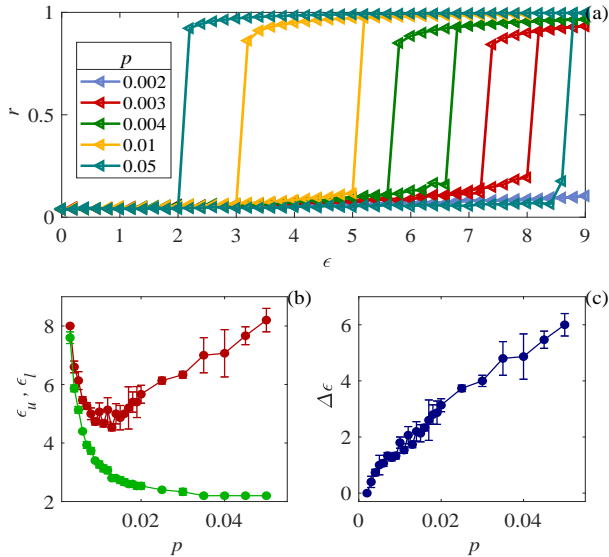}
\caption{(a) Results of ES in temporal networks emulated with configuration 3. (b) Critical values of the coupling strength required to synchronize ($\epsilon_u$: \mycircle{col1}), and critical values of coupling strength required to unsynchronize oscillators ($\epsilon_l$: \mycircle{col2}), (c) width of the hysteresis loops formed during ES.}
\label{6}
\end{figure}

\subsection{Configuration 3}

In the final scenario, the coupling is implemented such that at every iteration the oscillators randomly switch coupling between each other. This signifies that coupling between two oscillators does not depend on the physical distance between them and that the degree of the oscillators changes in every iteration. If the probability with which an oscillator couple to another while switching coupling is given by $p$ then the average degree of the network will be $k=pN$. Moreover, when an oscillator switches its coupling partners, the algorithm does not preclude that this oscillator cannot couple again to the same oscillators it was coupled in the previous iteration.

\cref{6} shows the results of ES as a function of $\epsilon$ for different values of $p$. Similar to previous results, for a sufficiently smaller $p$ (or $k$), ES is not observed in the population of oscillators and as $p$ increases, ES appears. In \cref{6}(b), variations of $\epsilon_u$ and $\epsilon_l$ are shown. Surprisingly, similar to the result for $v=0.05$ and $v=0.1$ in \cref{4}(b), $\epsilon_u$ in the present case, initially decreases with $p$ before starting to increase. On one hand, a possible explanation of this behavior can be ascribed to the fact that the oscillators in this case switch coupling at every iteration and the coupling also does not depend on the physical distance between the oscillators. This is tantamount to the situation that the oscillators, in this case, are moving with very high velocities and frequently changing their coupling partners. On the other hand, the reason for the similarity of these results only with the corresponding results of configuration 2 suggests that unlike configuration 1, in configuration 2 and 3, degrees of the oscillators do not remain constant. However, we modified configuration 1 by introducing fluctuations in the degrees of the oscillators, but the behavior of $\epsilon_u$ similar to that in \cref{4}(b) and \cref{6}(b) was not obtained. Therefore, the initial fall of $\epsilon_u$ with $k$ remains unclear. Finally, in \cref{6}(c), the variation of the width of hysteresis loops is shown as a function of $p$ and it increases monotonically with $p$.

\begin{figure}[ht!]
\includegraphics[width=9cm,height=4cm]{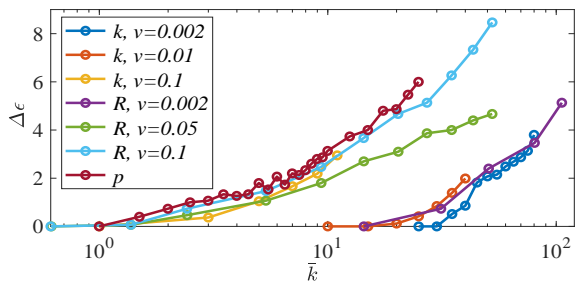}
\caption{Results of variation of $\Delta\epsilon$ with average degrees ($\bar{k}$) of the oscillators for three different configurations. }
\label{7}
\end{figure}

In \cref{7} the hysteresis widths ($\Delta\epsilon$) for all three configurations are plotted as a function of average degrees ($\bar{k}$); for configuration 1, $\bar{k}=k$, for configuration 2, $\bar{k}\propto R^2$, for configuration 3, $\bar{k}=pN$. It can be observed that for the situations when the velocities of the oscillators are relatively small ($v=0.002$ and $v=0.01$), the $\Delta\epsilon$ curves group together and for higher velocities or configuration 3 (which is identical to a case of higher velocity) $\Delta\epsilon$ curves formed another group.

\section{Conclusions}
We presented our results on ES in temporal networks of phase oscillators. The results were obtained and compared by implementing three different configurations in order to achieve the temporal nature of the networks. In two of these configurations, the oscillators were changing their coupling partners depending on the distances between them while in the third coupling partners switched randomly (irrespective of physical distances). Different control parameters were monitored in order to observe and analyze ES in the oscillators. Using \cref{eq6} the analytical understanding of the dependence of ES on $k$ and $v$ was established.

The most striking differences in three configurations were observed when degrees of the oscillators ($k\propto R^2$ in configuration 2 and $k\approx pN$ in configuration 3) were varied. In configuration 1 the critical value of coupling to observe ES ($\epsilon_u$) increased uniformly with $k$. However, in configuration 2 and 3, before starting to rise for larger values of $k$, $\epsilon_u$ decreased initially as $k$ started to increase from its lower values. Moreover, this behavior was observed only in situations when the oscillators were moving with large velocities. It was also observed that $\epsilon_l$ decreased and the width of the hysteresis loops increased with $k$ in all the configurations. The interaction time between the oscillators was also calculated using the ranges of different tie scales involved. The interaction time was calculated only for configuration 2 as the length scales were known. However, we expect that the interaction time between the oscillators modeled with configuration 1 will also be identical to that of configuration 2. The effect of density of the oscillators in observing ES can also be studied by varying the number of oscillators in the closed box. However, the effect of density can only be implemented for configuration 2 and we believe that results will be similar to those obtained by gradually varying $R$ (\cref{4}).

In our future work, we will explore ES by implementing the temporal nature of coupling in experimental systems. Configuration 1 and 2 although are more realistic to observe in nature, but are difficult to realize in model experimental nonlinear oscillators. Configuration 3, however, can be established easily and ES can be explored. The fact that ES has already been reported in an experimental system consisting of static Mercury Beating Heart oscillators \cite{Pawan} makes this system a potential candidate to study ES in the present circumstances.

\section{Data Availability}
We did not generate any data for our work. All results can be obtained by simulating the model represented by \cref{eq3} at different parameter values.
\bibliography{biblio}
\end{document}